\documentclass{article}

\usepackage[english]{babel}

\usepackage[letterpaper,top=2cm,bottom=2cm,left=3cm,right=3cm,marginparwidth=1.75cm]{geometry}

\usepackage{amsmath}
\usepackage{graphicx}
\usepackage[colorlinks=true, allcolors=blue]{hyperref}

\usepackage{times}
\usepackage{latexsym}
\usepackage{graphicx}
\usepackage{amsmath,amsthm,bm}
\usepackage{amsfonts}
\usepackage{booktabs}
\usepackage{multirow}
\usepackage{booktabs}
\usepackage{color}
\usepackage{CJKutf8}

\usepackage[T1]{fontenc}

\usepackage[utf8]{inputenc}

\title{Hybrid Retrieval and Multi-stage Text Ranking Solution at TREC 2022 Deep Learning Track}
\author{Guangwei Xu, Yangzhao Zhang, Longhui Zhang, Dingkun Long, Pengjun Xie, Ruijie Guo \\
  Alibaba Group \\
  \texttt{kunka.xgw,zhangyanzhao.zyz@alibaba-inc.com} \\
  \texttt{pinghe.zlh,dingkun.ldk@alibaba-inc.com} \\
  \texttt{pengjun.xpj,ruijie.guo@alibaba-inc.com} \\
}

\date{}

\begin{document}
\begin{CJK}{UTF8}{gbsn}
\maketitle

\begin{abstract}

Large-scale text retrieval technology has been widely used in various practical business scenarios. This paper presents our systems for the TREC 2022 Deep Learning Track. We explain the hybrid text retrieval and multi-stage text ranking method adopted in our solution. The retrieval stage combined the two structures of traditional sparse retrieval and neural dense retrieval. In the ranking stage, in addition to the full interaction-based ranking model built on large pre-trained language model, we also proposes a lightweight sub-ranking module to further enhance the final text ranking performance. Evaluation results demonstrate the effectiveness of our proposed approach. Our models achieve the {\bf 1st} and {\bf 4th} rank on the test set of passage ranking and document ranking respectively. 

\end{abstract}

\section{Introduction}


The text retrieval task is usually divided into two sub tasks according to the length of the text: passage retrieval and document retrieval. Whether it is a Document ranking or Passage ranking task, under the setting of full ranking, it generally needs to be processed in two stages: retrieval and ranking. Among them, the retrieval stage needs to quickly find the most relevant {\bf top-k} candidates in the entire corpus set given the input query, and then a more complex and accurate ranking stage will be performed over the {\bf top-k} candidates thus producing the final ranked result.

\noindent For the retrieval stage, BM25~\cite{robertson2009probabilistic} is a bag-of-words retrieval method that ranks a set of documents based on the query terms appearing in each document and is one of the best retrieval algorithms. Recently, various improved methods~\cite{nogueira2019doc2query,formal2021splade} based on BM25 have been proposed in the sparse retrieval research field. With the development of deep neural network models, the performance of dense retrieval models such as DPR~\cite{karpukhin-etal-2020-dense}, and coCondenser~\cite{Gao2021CondenserAP,Gao2022UnsupervisedCA} have surpassed traditional methods with a large margin. For the ranking stage, since the amount of processed data is greatly reduced, a more complex model structure can be adopted. Benefit from the excellent performance of large-scale pre-trained language model (e,g. BERT~\cite{devlin2018bert}) on various natural language processing tasks, the ranking model is also gradually turning to the BERT model as the relevance scoring model. Specifically, the query and doc will be concatenated as the input of BERT or other pre-trained language models like RoBerta~\cite{liu2019roberta}, ERNIE~\cite{lu2022erniesearch}, ELECTRA~\cite{clark2020electra}, 
 and Deberta~\cite{he2020deberta}. 


\noindent Based on this multi-stage processing method, our main optimization methods in this evaluation are:

1. In the retrieval stage, a hybrid retrieval method that combines traditional sparse retrieval and dense retrieval model is adopted, and the ROM~\cite{long2022retrieval} model we proposed for retrieval tasks is used on the dense retrieval pre-training model.

2. In the ranking stage, multiple backbone networks are used to train the ranking models, and the re-ranking HLATR~\cite{zhang2022hlatr} model we proposed before is used to further improve the final ensemble performance when merging these ranking models.

\section{Methodology}
\begin{figure}[h]
  \centering
  \includegraphics[width=0.8\linewidth]{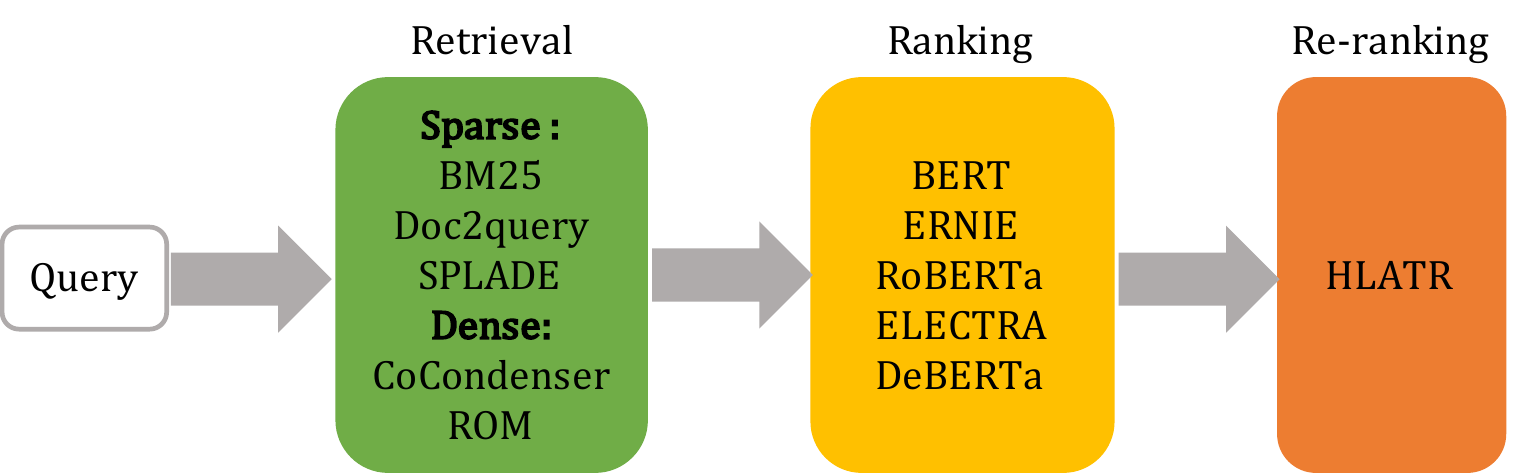}
  \caption{Model Architecture Diagram.}
  \label{fig:architecture}
\end{figure}

From the query to the final retrieval and ranking results, our method is divided into three stages: retrieval, ranking, and HLATR re-ranking. The specific process is shown in Figure~\ref{fig:architecture}. The difference between the Document ranking and the Passage ranking is that document is first divided into passages to participate in the retrieval and ranking, and finally the passage with the highest score is taken as the representative of the document. The following will introduce the specific practices of these three stages.

\subsection{Retrieval stage}

\paragraph{Sparse retrieval} BM25 has a good performance as a classic efficient scoring algorithm in the retrieval stage, but it is limited by the text matching between query and doc, and has no semantic expansion ability. Doc2query and SPLADE have extended the semantics from different perspectives, which can bring significant improvement in the sparse retrieval mode. Here we use BM25, Doc2query and SPLADE at the same time to perform weighted ensemble on the final score.

\paragraph{Dense retrieval} The retrieval stage is limited by computational efficiency. At present, the two-tower structure is basically used. The key is how to obtain a good twin-tower text semantic representation model. As a pre-training model, BERT can make full use of unlabeled corpus and can easily do fine-tuning downstream tasks, and is widely used here. For retrieval tasks, there are also many works to optimize the model structure or training loss function. The ROM model we use is optimized for the random mask in the MLM task to a weighted mask based on term weighting, which reduces the probability of the stop words that have a weak role in retrieval being masked. Such adjustments can make the final trained model more suitable for retrieval tasks.

\paragraph{Hybrid retrieval} Although the performance of the dense retrieval method has far exceeded that of the sparse retrieval method, we found that the results of the two methods can be further improved by weighted ensemble. It may be that the focus of the two methods is different.. The sparse retrieval method pays more attention to the text literal matching information, and the dense retrieval method pays more attention to the semantic matching information, and the two can complement each other.

\subsection{Ranking stage}

\paragraph{Single-tower interaction model} Compared with the retrieval stage, the amount of data processed in this stage is greatly reduced, and a single-tower model structure in which query and doc interact together can be used to model more accurately. The pre-training model of the BERT structure is also used here, and negative sampling is performed based on the results of the retrieval stage, and the rdrop method is used for the loss function.

\paragraph{Multi-model ensemble} Here we tried a variety of backbone models for training, including BERT, RoBERTa, ERNIE, ELECTRA, and DeBERTa. In the last model trained under different parameter settings, ten models with better performance were selected, while keeping the diversity of the backbone as much as possible.

\subsection{HLATR re-ranking stage}

\begin{figure}[t]
  \centering
  \includegraphics[width=0.6\linewidth]{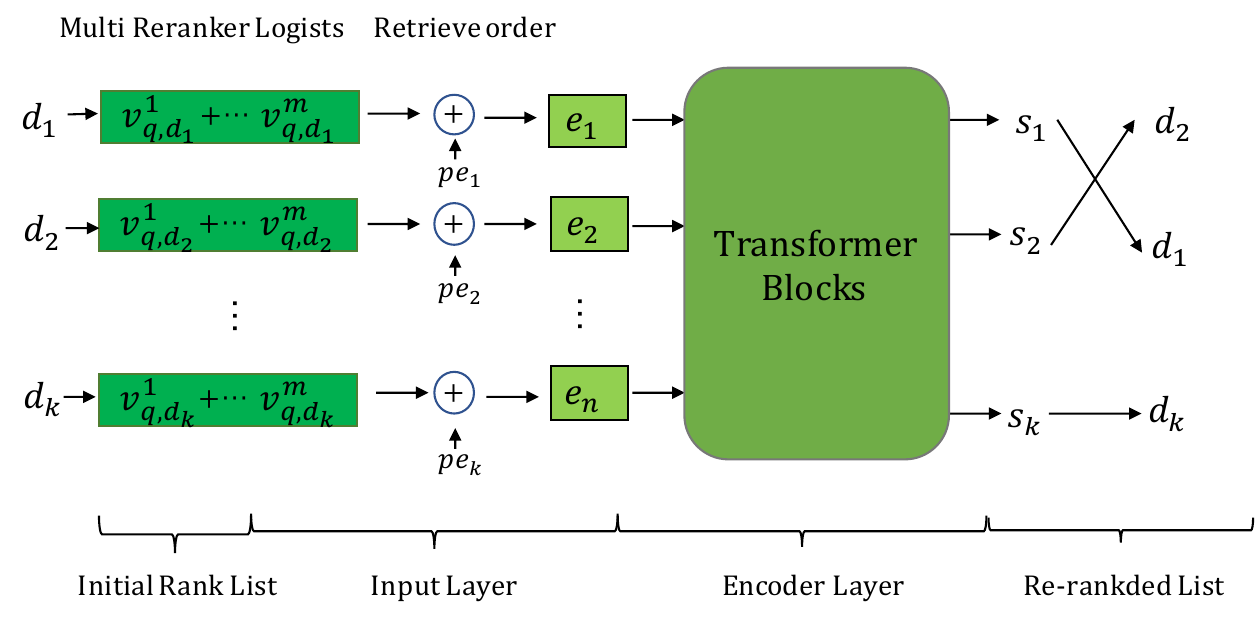}
  \caption{HLATR Re-ranking Model.}
  \label{fig:hlatr}
\end{figure}

When training the ranking model, we found that as the candidate set participating in the ranking stage increases, the final ranking performance will decrease. In the training stage, each query only interacts with a small number of negative samples, but in the testing stage, a larger candidate set needs to be interacted with. In addition, we made a simple weighted ensemble between the ranking model stage and the retrieval stage, and found that it can bring significant improvement. Based on these two findings, we introduce the HLATR re-ranking stage, using the Transformer structure to fuse the first two stages together. The model structure is shown in Figure~\ref{fig:hlatr}.

\section{Results}

\subsection{Main results}
The running results we finally submitted, the Passage ranking task is in Table~\ref{tab:result-passage}, and the Document ranking task is in Table~\ref{tab:result-document}.


\begin{table*}[t]
\centering
  \caption{Final results on TREC 2022 passage dataset.}
  \label{tab:result-passage}
  \begin{tabular}{ccc}
    \toprule
    Run & NDCG@10(Dev) & NDCG@10(Eval) \\
    \midrule
    pass1 & 0.7602 & 0.7050 \\
    pass2 & 0.7568 & 0.7105 \\
    pass3 & 0.7572 & {\bf 0.7184} \\
    \bottomrule
\end{tabular}
\end{table*}

\begin{table*}[t]
\centering
  \caption{Final results on TREC 2022 document dataset.}
  \label{tab:result-document}
  \begin{tabular}{ccc}
    \toprule
    Run & NDCG@10(Dev) & NDCG@10(Eval) \\
    \midrule
    doc1 & 0.7643 & 0.4908 \\
    doc2 & 0.7631 & 0.4555 \\
    doc3 & - & {\bf 0.7533} \\
    \bottomrule
\end{tabular}
\end{table*}

\noindent The three sets of results we submitted for the Passage ranking task were produced by different model ensemble, and finally pass3 got the best results on the test set.

\noindent For the Document ranking task, the first two sets of results we submitted tried to train the model directly based on the document data, and doc3 obtained the highest score after training the model based on the passage data as the document score. It may be due to the difference in the distribution of Dev and Eval data, the results of the first two groups dropped a lot, and only doc3 finally got normal results.

\subsection{Ablation study}
Since there are too many model combinations involved in the ranking model stage and the HLATR re-ranking stage, including a lot of parameter tuning work. Different data sampling and training result selection will have different performance, so the ablation experiments of these two stages are not listed here. In contrast, the retrieval stage is more valuable. We list the experimental results on the Passage ranking task in Table~\ref{tab:ablation-passage}, and the experimental results on the Document ranking task in Table~\ref{tab:ablation-document}.

\begin{table*}[!ht]
\centering
  \caption{Ablation experiments result on TREC 2021 passage dataset.}
  \label{tab:ablation-passage}
  \begin{tabular}{ccccc}
    \toprule
    Model & AP & NDCG@10 & R@1000 & MRR@100 \\
    \midrule
    BM25 & 0.0977 & 0.3977 & 0.5835 & 0.5303 \\
    Doc2query & 0.1358 & 0.4463 & 0.6147 & 0.5076 \\
    ROM & 0.2965 & 0.6479 & 0.8251 & 0.7865 \\
    coCondenser & 0.2857 & 0.6520 & 0.8095 & 0.7601 \\
    BERT-large-ROM & 0.2919 & 0.6471 & 0.8235 & 0.7799 \\
    SPLADE & 0.3329 & 0.6844 & {\bf 0.8588} & 0.8289 \\
    ensemble & {\bf 0.3792} & {\bf 0.7426} & 0.8185 & {\bf 0.8841} \\
    \bottomrule
\end{tabular}
\end{table*}

\begin{table*}[!ht]
\centering
  \caption{Ablation experiments result on TREC 2021 document dataset.}
  \label{tab:ablation-document}
  \begin{tabular}{cc}
    \toprule
    Model & NDCG@10 \\
    \midrule
    BM25 & 0.4370 \\
    Doc2query & 0.5703 \\
    Condenser & 0.5778 \\
    ROM & 0.6115 \\
    SPLADE & 0.6202 \\
    coROM (trained on document) & 0.5497 \\
    coCondenser (trained on document) & 0.5871 \\
    coROM (trained on passage) & 0.5815 \\
    ensemble & {\bf 0.7401} \\
    \bottomrule
\end{tabular}
\end{table*}

\noindent Whether it is Passage ranking task or Document ranking task, BM25 is a strong baseline, and the Doc2query can further strengthen the baseline. The condenser model based on the dense retrieval architecture is similar to the ROM model, and SPLADE performed very well on this evaluation data set. The results of the ensemble of multiple models in the final retrieval stage are very close to the final submitted results, which shows the importance of this stage.

\section{Conclusion}
In this TREC 2022 Deep Learning Tack evaluation, we participated in the full ranking setting of both the Passage ranking and Document ranking tasks. Through the technology of hybrid retrieval and multi-stage ranking model, the ensemble of multiple current mainstream algorithms has achieved relatively good results. But there is still a lot of room for improvement in the ranking stage and Document ranking task.

\clearpage
\bibliographystyle{alpha}
\bibliography{sample}

\end{CJK}

\end{document}